\documentclass[11pt,letterpaper]{article}
\usepackage[T1]{fontenc}
\usepackage[utf8]{inputenc}
\DeclareUnicodeCharacter{03B2}{\ensuremath{\beta}}
\DeclareUnicodeCharacter{03BA}{\ensuremath{\kappa}}
\DeclareUnicodeCharacter{2212}{\ensuremath{-}}
\DeclareUnicodeCharacter{00D7}{\ensuremath{\times}}
\DeclareUnicodeCharacter{00B5}{\ensuremath{\mu}}
\DeclareUnicodeCharacter{207B}{\textsuperscript{-}}
\DeclareUnicodeCharacter{2079}{\textsuperscript{9}}
\DeclareUnicodeCharacter{2077}{\textsuperscript{7}}
\DeclareUnicodeCharacter{2078}{\textsuperscript{8}}
\DeclareUnicodeCharacter{2076}{\textsuperscript{6}}
\DeclareUnicodeCharacter{00B9}{\textsuperscript{1}}
\DeclareUnicodeCharacter{2264}{\ensuremath{\leq}}
\DeclareUnicodeCharacter{2265}{\ensuremath{\geq}}
\usepackage[margin=1in]{geometry}
\usepackage{amsmath,amssymb,mathptmx}
\usepackage{graphicx,cite,booktabs,array,longtable,tabularx,calc}
\usepackage{float,placeins}
\usepackage[font=small,labelfont=bf]{caption}
\usepackage{microtype,xurl}
\usepackage[hidelinks]{hyperref}
\hypersetup{
  pdftitle={Integrated Stress Responses and ABCB11 Repression in Drug-Exposed Human Hepatocytes},
  pdfauthor={Pratyush Shukla},
  pdfsubject={Exploratory human hepatocyte reanalysis with specificity controls},
  pdfkeywords={ABCB11, bile salt export pump, integrated stress response, toxicogenomics}}

\providecommand{\citep}[1]{\cite{#1}}
\makeatletter
\providecommand{\bstctlcite}[1]{\if@filesw\immediate\write\@auxout{\string\citation{#1}}\fi}
\makeatother
\usepackage{parskip}
\title{Integrated Stress Responses and ABCB11 Repression in Drug-Exposed Human Hepatocytes}
\author{Pratyush Shukla\\[4pt]
\small UT Southwestern Medical Center\\[2pt]
\small\href{mailto:Pratyush.Shukla@UTSouthwestern.edu}{\nolinkurl{Pratyush.Shukla@UTSouthwestern.edu}}}
\date{11 October 2026}
\begin{document}
\maketitle
\bstctlcite{IEEEstyle:control}
\subsection*{Abstract}
Drug-safety information systems must distinguish direct bile salt export pump (BSEP) inhibition from transcriptional changes in hepatic transport programs. We investigated whether repression of \emph{ABCB11}, which encodes BSEP, tracks an integrated stress response (ISR) after accounting for dose and liver-identity proxies. We analyzed 197 human hepatocyte RNA-seq dose conditions for 50 compounds using compound-centered models and 2,000 compound-bootstrap resamples. The association remained negative after dose and 436-gene identity adjustment ($\beta=-0.288$; 95\% interval, $-0.416$ to $-0.127$), and after jointly adding global-expression, oxidative-stress and apoptosis scores ($\beta=-0.198$; $-0.342$ to $-0.039$). The adjusted partial correlation ($r=-0.422$) exceeded in absolute magnitude those obtained for all but 1 of 5,000 expression-matched and 16 of 5,000 expression- and variance-matched random six-gene sets. Several comparator transporters also declined, indicating that the association was not restricted to \emph{ABCB11}. Adding an interpolated source-screen viability proxy to all controls retained a negative estimate but an interval including zero ($\beta=-0.190$; $-0.335$ to $0.009$; 152 conditions, 39 compounds). Exposure-restricted uncertainty likewise depended on identity adjustment. Published modules from 240 external microarray conditions for 80 perturbagens showed consistent stress--metabolic covariation, without gene-level replication. A separate assay/FDA audit exposed sparse joint coverage and source-dependent associations. Together, these analyses link hepatocyte stress to \emph{ABCB11} repression and provide a biologically grounded approach to integrating biochemical, transcriptional and regulatory evidence for drug-safety assessment.
\par\medskip\noindent\textbf{Keywords:} ABCB11; bile salt export pump; integrated stress response; primary human hepatocytes; toxicogenomics; drug-safety evidence integration.
\hypertarget{introduction}{%
\section{Introduction}\label{introduction}}

\begin{figure}[htbp]
\hypertarget{fig:structure}{%
\centering
\includegraphics[width=\textwidth]{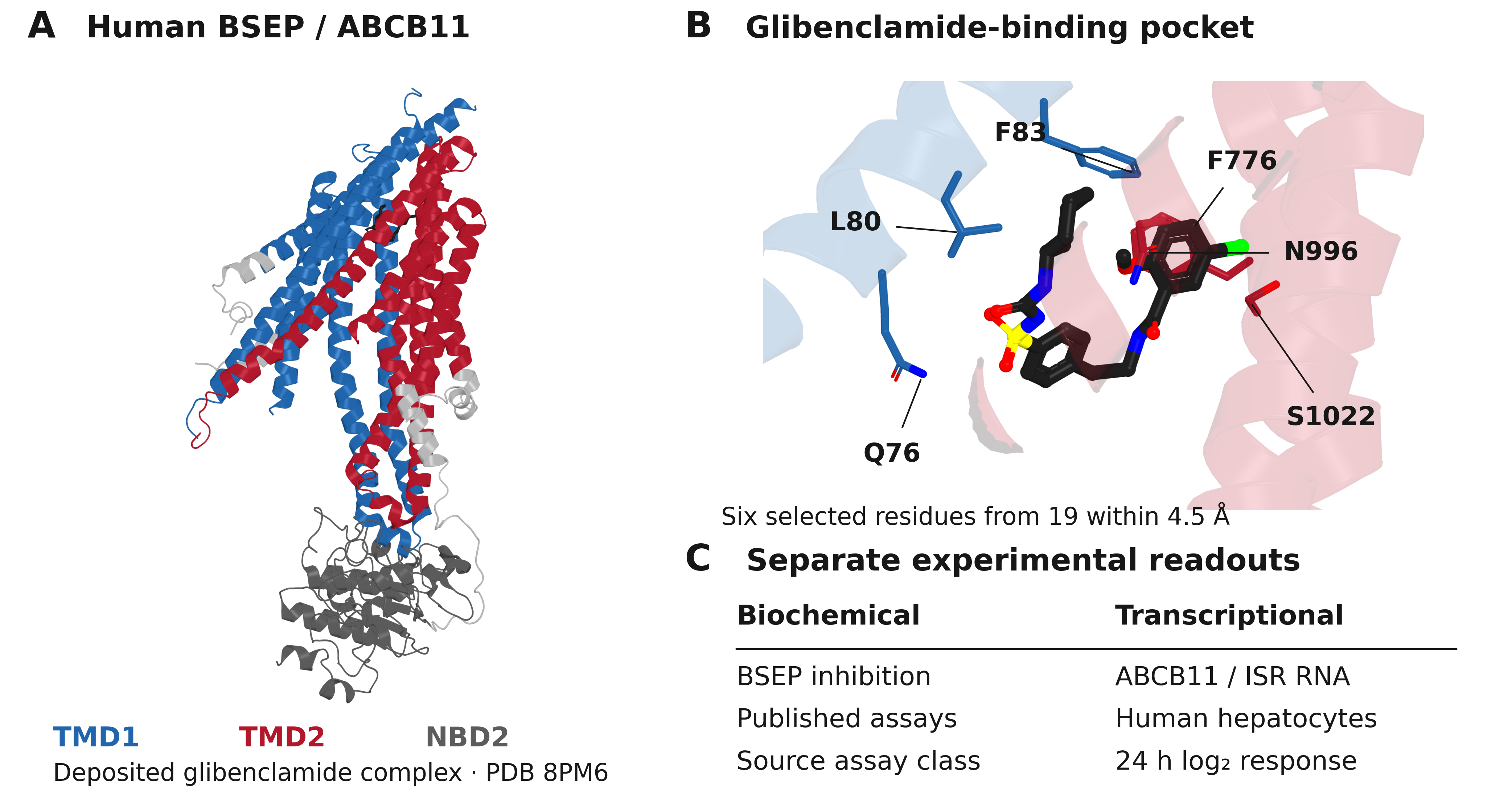}
\caption{Experimental human BSEP--glibenclamide structure and biological readouts. A: TMD1 blue, TMD2 red and resolved NBD2 gray. B: deposited ligand and six selected residues among 19 within 4.5 Å; ligand carbons are charcoal, heteroatoms colored by element. C: separate biochemical and transcriptional systems. NBD1 is unresolved in 8PM6. Coordinates: Liu and colleagues; original renderings.}\label{fig:structure}
}
\end{figure}

Drug-induced liver injury (DILI) emerges from interacting drug exposures, cellular liabilities, and host susceptibility. Biochemical BSEP inhibition and mitochondrial dysfunction have been associated with severe DILI in selected compound panels \citep{aleo}. BSEP, encoded by \emph{ABCB11}, exports bile salts across the hepatocyte canalicular membrane. A deposited human BSEP--glibenclamide structure provides a molecular account of direct inhibitor binding \citep{structure}. Transcriptional loss of transport capacity poses a different biological question: a drug may alter the expression program that sustains canalicular function without satisfying a particular biochemical inhibitor threshold. Figure~\ref{fig:structure} distinguishes the deposited protein structure and biochemical readout from the transcriptional responses analyzed here.

Stress-associated loss of hepatic identity already has substantial biological precedent. Dubois and colleagues demonstrated active repression of liver-identity programs during endoplasmic-reticulum stress, involving altered transcription-factor and enhancer activity \citep{identity}. ISR-related CHOP expression has also been linked to mitochondrial toxicants and DILI liability \citep{isr}. These observations motivate a specificity question: does \emph{ABCB11} repression merely accompany general hepatocyte identity loss, or does a residual association with stress remain after accounting for that broader response? A transcriptional ISR signature cannot by itself identify the initiating organelle or prove mitochondrial damage.

Drug-safety information systems need to distinguish a measured transport-inhibition result from an altered expression program and a regulatory concern label. Treating these observations as interchangeable can conceal missing evidence and overstate patient-risk inference. Recent human toxicogenomic resources make this biological distinction accessible to secondary analysis. DILImap supplies dose-resolved primary human hepatocyte RNA-seq responses \citep{dilimap}, while the human hepatocyte TXG-MAPr provides published co-expression modules derived from Open TG-GATEs \citep{txg}. FDA DILIrank 2.0 supplies updated regulatory and literature concern categories \citep{dilirank}. These resources measure different layers of evidence and have different compound ascertainment, assay coverage, and exposure scales. Their integration therefore requires preserving missingness and avoiding conversion of benchmark categories into patient-level risk.

We test whether stress-associated \emph{ABCB11} repression remains after accounting for dose, liver-identity proxies, broader expression changes and alternative stress programs. Matched random gene sets and comparator transporters test the scope of that association. External published modules provide a pathway-level consistency check. A separate assay/FDA audit illustrates why drug-safety evidence integration must preserve each source, exposure, readout and uncertainty, rather than infer clinical safety from a missing or thresholded measurement.

\hypertarget{materials-and-methods}{%
\section{Materials and methods}\label{materials-and-methods}}

\hypertarget{study-design-and-analysis-sequence}{%
\subsection{Study design and analysis sequence}\label{study-design-and-analysis-sequence}}

This study reanalyzed public data and generated no new wet-laboratory or clinical measurements. An initial local specification defined four measured-assay comparisons against FDA concern categories. After that analysis exposed sparse joint coverage, a separate transcriptomic specification was recorded before examining response associations. The entire transcriptomic extension is exploratory. Broad-identity, external-module, source-unit and the additional specificity controls described below were later exploratory amendments. The study was not preregistered.

\hypertarget{human-hepatocyte-rna-seq-and-exposure-audit}{%
\subsection{Human hepatocyte RNA-seq and exposure audit}\label{human-hepatocyte-rna-seq-and-exposure-audit}}

We downloaded the publicly served DILImap validation DESeq2 and raw-count files on 10 October 2026. This public snapshot contains 208 response conditions for 51 compounds, extending beyond the source article's original clinical validation panel. All response conditions were at 24 hours. We used the authors' signed log2 fold changes against source controls and retained source false-discovery rates (FDRs), including missing values, without selecting genes by statistical significance. DESeq2 supplies the underlying differential-expression framework \citep{deseq}.

Exact compound and compound-dose-plate identifiers linked every response condition to sample quality-control metadata. A high-dose lubiprostone condition was conservatively excluded because a matched sample carried a cell-death flag. Ten repeated chlorpromazine reference conditions, all at one concentration, were excluded from dose-series inference. The retained panel comprised 197 conditions from 50 compounds, each with at least three source concentrations. Biological variation was analyzed at the compound level; replicate wells were not treated as independent donors.

Nominal concentrations were divided by source median total plasma Cmax where available. A sensitivity retained conditions at or below 10× Cmax and required at least two remaining doses per compound, yielding 80 conditions from 27 compounds. Another sensitivity used the source maximum tested dose designated for the IC10 criterion, yielding 177 conditions from 45 compounds with repeated eligible doses. These restrictions contextualize exposure; nominal culture concentration and total plasma Cmax do not establish free intracellular exposure or clinical relevance. The source PHH model was based on one donor, limiting donor-generalization \citep{dilimap}.

\hypertarget{fixed-biological-scores-and-identity-comparators}{%
\subsection{Fixed biological scores and identity comparators}\label{fixed-biological-scores-and-identity-comparators}}

The ISR score was the mean signed log2 fold change of \emph{DDIT3, ATF3, ASNS, TRIB3, CHAC1}, and \emph{PPP1R15A}. The individual transcript endpoint was \emph{ABCB11}. A canalicular score averaged \emph{ABCB11, ABCB4}, and \emph{ABCC2}; these transporters handle different substrates, so the composite is not exclusively a bile-acid export score. All fixed-score genes had finite responses across source conditions.

The initial identity comparator averaged \emph{HNF4A, ALB, TTR}, and \emph{APOA1}. A subsequent specificity check used the independently published mouse LIVER-ID panel \citep{identity}. Of 621 source symbols, 455 matched human RNA-seq symbols after case normalization. After removing overlapping endpoint/ISR genes and genes with any nonfinite source response, a uniform 436-gene panel remained. This is an exact-symbol projection of a published identity panel, with an explicit mapping audit, rather than a complete curated orthology map. A similarly processed, disjoint 443-gene ubiquitous-expression panel provided descriptive context. Genes were not chosen by their associations with \emph{ABCB11} or the ISR.

\hypertarget{external-human-hepatocyte-module-analysis}{%
\subsection{External human hepatocyte module analysis}\label{external-human-hepatocyte-module-analysis}}

We used TXG-MAPr's openly published module eigengene scores and memberships, derived from primary human hepatocyte Open TG-GATEs microarrays \citep{txg, tggates}. Membership inspection identified PHH:15 as an ATF4-related module containing \emph{ASNS, TRIB3}, and \emph{CHAC1}, and PHH:2 as a 743-gene metabolic/mitochondrial module containing \emph{ABCB11} and \emph{ABCC2}. Both transport genes have positive gene--module correlations, approximately 0.409 and 0.370. PHH:2 is a broad program and cannot isolate either transcript's response.

The published TXG-MAPr supplementary dataset contained 941 conditions across 158 perturbagens. Restricting to 24 hours and complete three-level dose series retained 240 conditions from 80 chemical and biological perturbagens. A disjoint identity comparator averaged PHH:9, containing \emph{TTR} and \emph{APOA1}, and PHH:218, containing \emph{HNF4A}. Source dose units were constant within each perturbagen but heterogeneous across perturbagens; therefore dose adjustment used within-perturbagen log source dose. A later sensitivity retained the 69 perturbagens explicitly annotated in µM, avoiding interpretation of questionable source unit annotations. The external analysis tests consistency of fixed published module responses; the source authors learned those modules from TG-GATEs itself.

\hypertarget{statistical-estimation}{%
\subsection{Statistical estimation}\label{statistical-estimation}}

For each transcript endpoint, we fit models of the form

\[Y_{dc} = \alpha_d + \beta S_{dc} + \gamma\log_2 C_{dc} + \delta I_{dc} + \varepsilon_{dc},\]

where \(d\) denotes a compound, \(c\) a concentration, \(S\) the ISR score, and \(I\) an identity comparator. Variants omitted dose, identity, or both. Within-compound centering removed compound intercepts. We estimated 95\% percentile intervals with 2,000 compound-cluster bootstrap resamples, seed 20261010. Cluster-robust t-tests used compound clusters and finite-sample corrections counting absorbed compound intercepts. Holm adjustment covered the two unadjusted transcript endpoints; additional checks remained explicitly exploratory. We calculated every leave-one-compound-out slope and retained the full gene-level estimates. Identity-member outcomes were not adjusted by a composite containing themselves.

The external models used the same estimation procedure on module scores. Regression coefficients are observational associations. Adjustment can neither establish mediation nor identify a causal effect, and coefficients on RNA log2 fold changes and normalized module eigengenes are not directly equivalent biological effect sizes.

\hypertarget{additional-specificity-controls}{%
\subsection{Additional exploratory specificity controls}\label{additional-specificity-controls}}

We added mean, median and mean absolute log2 fold-change controls from 16,133 genes with finite responses in all source conditions and positive mean vehicle counts per million (CPM). ISR, endpoint, comparator-outcome and broad-identity genes were excluded. Baseline expression used 80 DMSO wells. Separate scores averaged 48 reactive-oxygen-species (ROS), 145 apoptosis and 105 ISR-disjoint unfolded-protein-response genes from MSigDB Hallmarks \citep{hallmark}, excluding unavailable/nonfinite genes and overlap with named ISR/endpoints. These expression scores do not measure ROS, apoptosis or absolute RNA loss.

For a random-set benchmark, we generated 5,000 six-gene sets matched to each ISR member's baseline-expression decile, and 5,000 additionally matched to its within-compound response-variance quintile (seed 20261011). Sets sampled distinct genes from the 16,133-gene background. After compound centering and dose/identity adjustment, we compared absolute partial correlations with \emph{ABCB11}; empirical exceedance fractions were $(1+k)/5001$. This benchmark tests unusual covariation relative to matched random genes, not ISR exclusivity among biological pathways. Eleven transporter/housekeeping outcomes used the same adjusted model, removing any outcome from the identity comparator that contained it; exploratory BH correction covered these outcomes.

Source-screen LDH viability was normalized to mean DMSO viability on each screening plate and averaged by exact compound name/dose. Only 38 retained RNA conditions had exact dose matches, with repeated matches for one compound, precluding a within-compound viability model. A sensitivity therefore interpolated viability linearly in log10 dose within each source range, without extrapolation, covering 152 conditions from 39 compounds. We compared models before/after this proxy adjustment on the same panel and repeated the adjusted model below the source IC10-designated dose ceiling. Finally, joint models combined global-expression and alternative-stress controls, with/without the LDH proxy. All additional controls were post hoc and used pointwise bootstrap intervals.

\hypertarget{measured-assays-and-fda-concern-annotations}{%
\subsection{Measured assays and FDA concern annotations}\label{measured-assays-and-fda-concern-annotations}}

We extracted experimental BSEP and mitochondrial-toxicity annotations from the original-data resource accompanying Seal and colleagues \citep{seal}, tracing those panels to McLoughlin and colleagues and Hemmerich and colleagues \citep{bsep, mito}. Only measured source rows were used; pretrained toxicity predictions were excluded. A second published BSEP compilation supplied a source-sensitivity panel \citep{bsep2}. Its inhibitor classification used IC50 below 25 µM, with non-inhibitors above 100 µM and an intermediate exclusion region. Thus a source-negative label is a thresholded assay annotation, not proof of absent interaction at every concentration.

DrugCentral structures \citep{drugcentral} and successful exact-name PubChem retrievals supported parent-level joins. Exact complete names were preferred, allowing a finite counterion-normalization list. RDKit removed salts, neutralized structures, canonicalized tautomers and removed stereochemistry. Conflicting identities, assay labels and FDA groups were excluded. This harmonization can obscure stereospecific biology; RNA models retained full compound identities.

From 1,336 FDA records, we retained 953 unique parents: 175 most-, 280 less-, 238 no- and 260 ambiguous-concern. Four initial BSEP/mitochondrial comparisons contrasted most versus no and most versus less concern, using conditional odds ratios, exact 95\% intervals, Fisher tests and Holm correction. Unmeasured assays remained missing. Joint prediction required at least 50 complete parents and 10 per class; only 40 most/no parents were complete, so cross-validation was not performed. DILIrank concern categories are not patient incidence or safe-exposure thresholds \citep{dilirank}.

\hypertarget{structural-figures-and-provenance}{%
\subsection{Structural figures and provenance}\label{structural-figures-and-provenance}}

Protein images were rendered directly from experimental human BSEP--glibenclamide coordinates, PDB 8PM6 \citep{structure}. Domain boundaries followed UniProt annotations. NBD1 was unresolved in this deposited structure and was not reconstructed. Nineteen residues have deposited heavy atoms within 4.5 Å of the ligand; six are illustrated in Figure~\ref{fig:structure}. These distances describe existing coordinates; no docking, new affinity measurement, or new structure determination was performed. Quantitative charts were generated from the analysis results; protein visualizations were rendered from deposited experimental coordinates.

\hypertarget{results}{%
\section{Results}\label{results}}

\hypertarget{stress-associated-abcb11-repression-persists-after-identity-adjustment}{%
\subsection{Stress-associated ABCB11 repression persists after identity adjustment}\label{stress-associated-abcb11-repression-persists-after-identity-adjustment}}

Across 197 retained conditions, the within-compound \emph{ABCB11} slope per ISR unit was −0.673 (95\% compound-bootstrap interval, −0.829 to −0.524). The canalicular composite slope was −0.353 (−0.453 to −0.261). Holm-adjusted cluster p-values were 1.97×10⁻⁹ and 1.13×10⁻⁷, respectively. These relationships concern concentration-dependent responses within compounds rather than a comparison of unrelated high- and low-concern drug classes.

\begin{figure}[htbp]
\hypertarget{fig:rna}{%
\centering
\includegraphics[width=\textwidth]{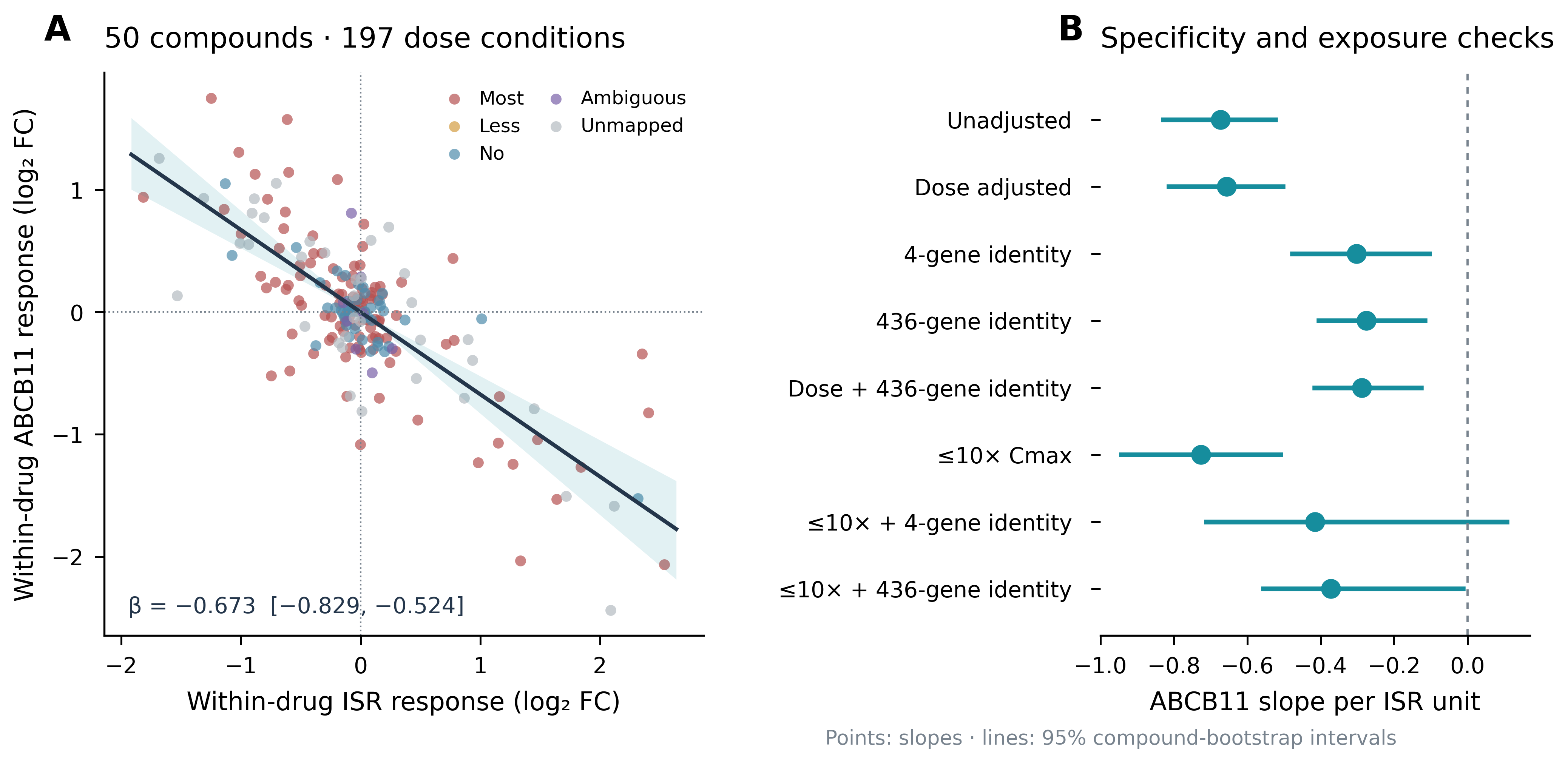}
\caption{Within-compound RNA-seq responses and specificity checks. Each point is a retained drug--dose condition; colors indicate matched FDA concern categories, with gray for unmapped parents. The fitted line and shaded band use the unadjusted ISR slope and its compound-bootstrap interval. Forest estimates show ABCB11 slopes and 95\% compound-bootstrap intervals; the low-exposure fits use 80 conditions from 27 compounds. The four-gene identity-adjusted low-exposure interval crosses zero.}\label{fig:rna}
}
\end{figure}

Dose adjustment changed the \emph{ABCB11} slope little, to −0.657 (−0.813 to −0.503). Adjustment by the four-gene identity score attenuated it to −0.302 (−0.477 to −0.103). Using the broader 436-gene comparator gave −0.276 (−0.405 to −0.116); jointly adjusting for dose and that comparator gave −0.288 (−0.416 to −0.127). The fully adjusted canalicular score remained inversely associated with ISR, β = −0.153 (−0.252 to −0.077). Figure~\ref{fig:rna} shows the compound-centered association and principal identity/exposure checks.

No single compound explained the \emph{ABCB11} direction: unadjusted leave-one-compound-out slopes ranged from −0.717 to −0.640, and slopes after dose plus broad-identity adjustment ranged from −0.332 to −0.259. The broad identity score itself declined with ISR, β = −0.093 (−0.123 to −0.067), while the ubiquitous comparator had β = 0.019 (−0.005 to 0.042). These descriptive contrasts contextualize \emph{ABCB11} repression but do not prove a transporter-specific regulatory mechanism.

\hypertarget{additional-controls-refine-the-scope-of-the-association}{%
\subsection{Additional controls refine the scope of the association}\label{additional-controls-refine-the-scope-of-the-association}}

After dose/broad-identity adjustment, adding mean, median or absolute global-expression changes separately gave \emph{ABCB11} slopes of $-0.290$, $-0.231$ and $-0.274$, respectively; all pointwise bootstrap intervals excluded zero. Joint median/absolute adjustment gave $-0.263$ ($-0.411$ to $-0.074$). Adding ROS and apoptosis scores jointly gave $-0.227$ ($-0.358$ to $-0.105$); the ISR-disjoint unfolded-protein-response adjustment gave $-0.324$ ($-0.555$ to $-0.091$). Combining global-expression and ROS/apoptosis controls attenuated the slope to $-0.198$ ($-0.342$ to $-0.039$).

\begin{figure}[htbp]
\centering
\includegraphics[width=\textwidth]{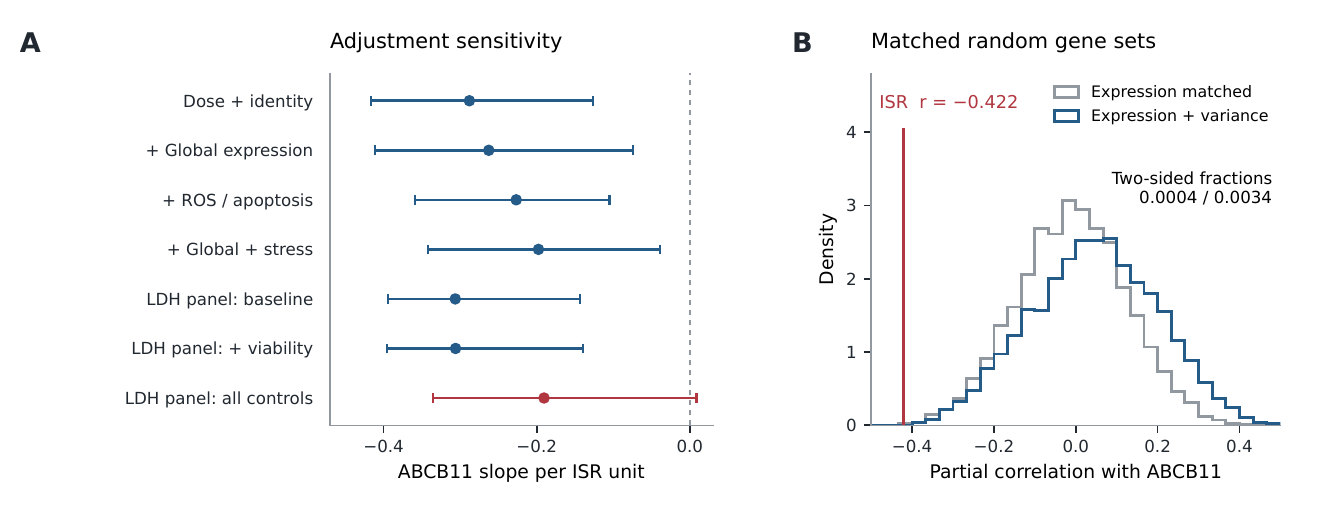}
\caption{Additional exploratory specificity controls. A: ABCB11 slopes with 95\% pointwise compound-bootstrap intervals; every model adjusts for dose and broad identity. Global controls combine median and absolute expression change; stress controls combine ROS/apoptosis scores. LDH models use 152 conditions/39 compounds; the other displayed models use 197/50. The final interval crosses zero. B: partial-correlation distributions for 5,000 random six-gene sets per benchmark after dose/identity adjustment. The red line marks the fixed ISR score; displayed fractions count both tails. Matching assesses unusual covariation, not causality or pathway exclusivity.}\label{fig:controls}
\end{figure}

The dose/identity-adjusted partial correlation was $r=-0.422$. Only 1/5,000 expression-matched and 16/5,000 expression/variance-matched random sets had equally large absolute correlations, giving empirical fractions of 0.0004 and 0.0034 (Figure~\ref{fig:controls}). These comparisons support an unusual association relative to the sampled background, rather than a unique ISR mechanism.

In the LDH-coverage panel, proxy adjustment changed the slope little, from $-0.306$ ($-0.394$ to $-0.144$) to $-0.306$ ($-0.396$ to $-0.139$). Below the IC10-designated dose ceiling, dose/identity adjustment gave $-0.292$ ($-0.423$ to $-0.128$; 177 conditions, 45 compounds). However, jointly adding the LDH proxy to all global-expression and ROS/apoptosis controls gave $-0.190$ ($-0.335$ to $0.009$); this most extensively adjusted interval included zero.

Comparator outcomes constrain transporter specificity (Table~\ref{tab:comparators}). Several transporters remained negatively associated with ISR; \emph{ABCB4}, \emph{ABCC4} and \emph{SLC51B} intervals included zero. All three housekeeping intervals included zero. These comparator results indicate that the association is not restricted to \emph{ABCB11}.

\begin{table}[htbp]
\caption{Exploratory comparator outcomes}\label{tab:comparators}
\centering
\begin{tabular}{lrr}
\hline
Outcome & ISR slope & 95\% interval \\
\hline
\emph{ABCB4} & $-0.036$ & $[-0.257,\ 0.126]$ \\
\emph{ABCC2} & $-0.136$ & $[-0.221,\ -0.040]$ \\
\emph{SLC10A1} & $-0.313$ & $[-0.396,\ -0.224]$ \\
\emph{SLCO1B1} & $-0.117$ & $[-0.191,\ -0.028]$ \\
\emph{ABCC3} & $-0.122$ & $[-0.169,\ -0.070]$ \\
\emph{ABCC4} & $-0.134$ & $[-0.320,\ 0.094]$ \\
\emph{SLC51A} & $-0.293$ & $[-0.446,\ -0.151]$ \\
\emph{SLC51B} & $-0.066$ & $[-0.318,\ 0.185]$ \\
\emph{GAPDH} & $-0.016$ & $[-0.114,\ 0.070]$ \\
\emph{ACTB} & $-0.127$ & $[-0.231,\ 0.001]$ \\
\emph{RPLP0} & $0.051$ & $[-0.012,\ 0.099]$ \\
\hline
\end{tabular}

\smallskip
\parbox{\columnwidth}{\footnotesize All fits use 197 conditions/50 compounds, dose adjustment and broad identity with the outcome removed where necessary. Intervals are pointwise compound-bootstrap intervals. These are biological comparators, not validated negative controls.}
\end{table}

\hypertarget{exposure-restriction-retains-the-association-with-adjustment-dependent-uncertainty}{%
\subsection{Exposure restriction retains the association with adjustment-dependent uncertainty}\label{exposure-restriction-retains-the-association-with-adjustment-dependent-uncertainty}}

At or below 10× source total plasma Cmax, the unadjusted \emph{ABCB11} slope was −0.726 (−0.943 to −0.509). The four-gene identity-adjusted estimate was −0.416 (−0.711 to 0.108); its bootstrap interval included zero. Broad-identity adjustment gave −0.372 (−0.556 to −0.011), with a bound close to zero. The restricted canalicular estimates were likewise sensitive to the comparator and uncertainty method. These results support an exposure-restricted association while leaving its degree of specificity uncertain. Restriction to the source IC10-designated dose ceiling produced a similar unadjusted \emph{ABCB11} slope, −0.676 (−0.843 to −0.514).

\hypertarget{an-external-atf4metabolic-module-relationship-is-consistent-in-direction}{%
\subsection{An external ATF4--metabolic module relationship is consistent in direction}\label{an-external-atf4metabolic-module-relationship-is-consistent-in-direction}}

In the 80-perturbagen, 240-condition TG-GATEs panel, PHH:2 responses were inversely associated with PHH:15 responses, β = −0.660 (−0.799 to −0.543). Dose adjustment gave −0.489 (−0.626 to −0.357), and joint dose/identity adjustment gave −0.476 (−0.621 to −0.351). Every fully adjusted leave-one-perturbagen-out estimate remained negative, ranging from −0.506 to −0.455. Restriction to the 69 µM-annotated perturbagens gave a fully adjusted estimate of −0.507 (−0.654 to −0.380).

\begin{figure}[htbp]
\hypertarget{fig:external}{%
\centering
\includegraphics[width=\textwidth]{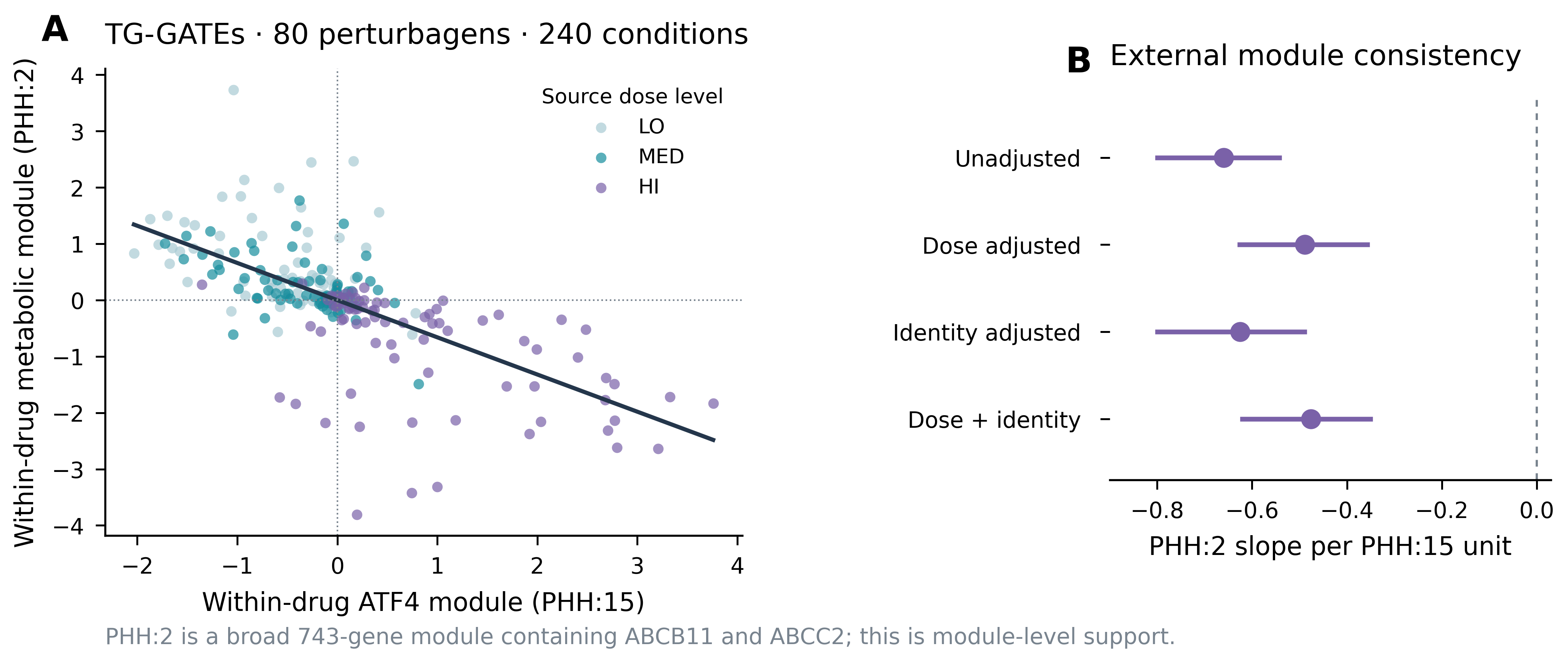}
\caption{External primary human hepatocyte module responses at 24 hours. Points are source dose conditions from 80 perturbagens, including chemical and biological exposures. The ATF4-related PHH:15 module is plotted against the broad PHH:2 metabolic/mitochondrial module, whose members include ABCB11 and ABCC2. Forest intervals use compound-cluster bootstrap resampling. The outcome is a published module eigengene, so the slopes are not directly interchangeable with RNA-seq transcript slopes.}\label{fig:external}
}
\end{figure}

The external pattern supports stress-associated loss of a broad metabolic program containing canalicular transport genes (Figure~\ref{fig:external}). The source module's 743 genes and modest \emph{ABCB11} module correlation limit gene-specific interpretation. Because individual external \emph{ABCB11} responses were not analyzed, direct gene-level replication remains outstanding.

\hypertarget{transcriptional-responses-cross-biochemical-threshold-categories}{%
\subsection{Transcriptional responses cross biochemical threshold categories}\label{transcriptional-responses-cross-biochemical-threshold-categories}}

All 50 RNA compounds had unambiguous exact-name source structures after standardization; 41 linked to current FDA categories. Assay coverage was limited: four linked to the primary BSEP panel, 24 to the secondary BSEP panel, and 13 to the mitochondrial panel. The secondary BSEP overlap contained three source-active and 21 source-negative compounds. At the highest retained RNA dose, eight of the 21 source-negative compounds had \emph{ABCB11} log2 fold change at or below −1. An exploratory active-versus-negative comparison was inconclusive (Mann--Whitney p = 0.680 for \emph{ABCB11}), with only three actives.

\begin{figure}[htbp]
\hypertarget{fig:cases}{%
\centering
\includegraphics[width=\textwidth]{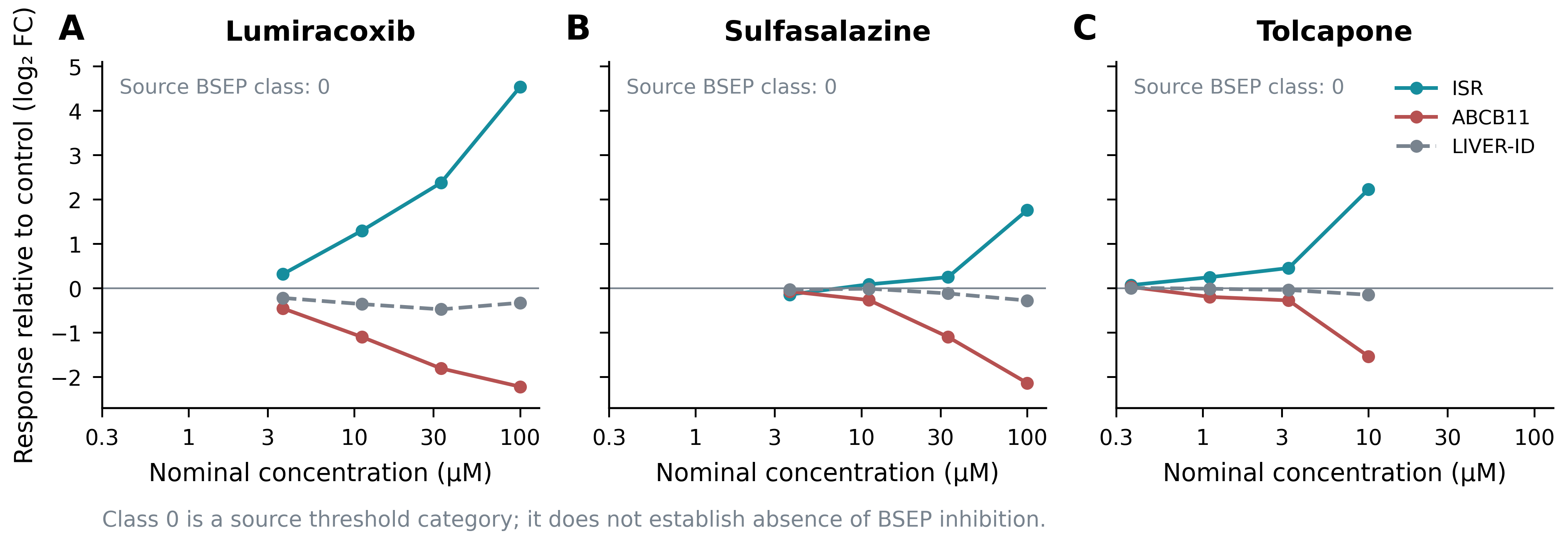}
\caption{Illustrative dose responses for three source-negative BSEP classifications. Curves show ISR, ABCB11, and the broad LIVER-ID comparator against nominal culture concentration. These descriptive cases were selected after analysis and are not a representative sample. Class 0 follows the published compilation threshold; tolcapone has reported BSEP interactions. Transcript repression and direct transport inhibition may coexist, and transporter protein abundance or flux was not measured here.}\label{fig:cases}
}
\end{figure}

Lumiracoxib, sulfasalazine, and tolcapone illustrate the interpretation boundary. Each had source BSEP class 0 yet an \emph{ABCB11} decrease exceeding twofold at a nominal concentration no higher than 100 µM and no higher than 10× source Cmax. At 33.3 µM lumiracoxib, \emph{ABCB11} log2 fold change was −1.808 (source FDR 1.63×10⁻¹⁸), versus −1.099 for sulfasalazine at 33.3 µM (3.03×10⁻⁷) and −1.538 for tolcapone at 10 µM (2.67×10⁻⁶). These cases were selected descriptively after complete-panel analysis (Figure~\ref{fig:cases}). Published mechanistic work reports BSEP inhibition and mitochondrial effects for tolcapone \citep{tolcapone}; its class 0 here underscores threshold dependence rather than establishing a transport-inhibition-free mechanism.

\hypertarget{sparse-and-source-dependent-assay-evidence-constrains-clinical-category-inference}{%
\subsection{Sparse and source-dependent assay evidence constrains clinical-category inference}\label{sparse-and-source-dependent-assay-evidence-constrains-clinical-category-inference}}

Among the 953 curated FDA parents, 200 had primary BSEP measurements, 313 mitochondrial annotations, and 114 both. In the initial most/no comparison, BSEP activity was 15/35 versus 13/38, conditional odds ratio 1.435 (exact interval, 0.504--4.140). Mitochondrial activity was 9/59 versus 3/62, odds ratio 3.505 (0.817--21.212). None of the four initial comparisons survived Holm correction; the minimum adjusted p-value was 0.283. Sparse data and wide intervals limit interpretation of these results.

The secondary BSEP panel gave an exploratory most/no odds ratio of 4.430 (1.528--15.799), despite 94.7\% agreement across 170 overlapping FDA parents. On the same 62 most/no parents, estimates were 2.043 (0.592--7.508) and 2.947 (0.703--14.944). The differing coverage, rather than a demonstrated difference between source effects, limits clinical-category inference. This audit motivates preserving assay thresholds and missingness in evidence integration; it does not validate a prediction model.

\FloatBarrier

\hypertarget{discussion}{%
\section{Discussion}\label{discussion}}

The central observation is a within-compound association between ISR induction and \emph{ABCB11} repression after adjustment for dose and two liver-identity proxies. Global-expression and alternative-stress controls attenuate but do not individually remove the association; matched random sets rarely reach the same absolute partial correlation. Several comparator transporters also decline, indicating that the observed association is not restricted to \emph{ABCB11}. However, the \emph{ACTB} slope is comparable to those of several transporter genes, limiting a transporter-selective interpretation. The most extensively adjusted model, including an interpolated LDH proxy, retains a negative estimate but an interval crossing zero.

The biological contribution builds on established stress-driven hepatic dedifferentiation \citep{identity} and multi-liability DILI biology \citep{aleo,isr}. It quantifies a residual stress--transport relationship and its sensitivity to competing explanations. Adjustment cannot establish that ISR causes repression or distinguish shared upstream regulation from mediation. Nor can mRNA establish BSEP protein abundance, localization, transport flux or a drug's clinical injury mechanism.

For a drug-safety information system, the practical distinction is simple: an assay-negative annotation describes one experimental threshold; reduced \emph{ABCB11} expression describes a cellular response; an FDA concern label summarizes clinical/regulatory evidence. None substitutes for the others. Their integration should preserve the compound identity, biological endpoint, exposure, source, missingness and uncertainty. The assay audit is useful precisely because sparse coverage and source selection prevent a credible joint predictive claim.

Important limitations remain. One RNA-seq donor cannot establish donor generalization. Both ISR and comparator scores are relative expression measurements; global-shift controls cannot rule out absolute RNA loss or normalization effects. The broad identity comparator is a mouse-panel exact-symbol projection, not a comprehensive human identity model. The source LDH curves are parallel screening data, and most usable doses require interpolation; they do not measure cytotoxicity in the RNA wells. Correlated stress controls may also adjust away part of the biology of interest. All added specificity checks are post hoc, and bootstrap intervals are pointwise. The restricted-exposure results remain dependent on identity definition; we prioritize the compound-bootstrap intervals.

External published modules add platform-level consistency, not held-out gene-level replication: the broad 743-gene metabolic module was learned from the source microarrays. The all-80 panel includes non-drug perturbagens, and the 69-perturbagen \ensuremath{\mu}M sensitivity does not solve intracellular exposure uncertainty. Assay heterogeneity, intermediate-inhibitor exclusion and stereochemistry removal also limit the separate biochemical/FDA audit. Regulatory categories are not patient outcomes measured in these cultures.

A discriminating next experiment would jointly measure \emph{ABCB11} mRNA, BSEP protein/localization, bile-salt flux and viability across multiple donors, doses and time points, with a stress-pathway perturbation. Direct BSEP inhibitors, the illustrative drugs and exposure-matched low-response compounds would help distinguish adaptive transport remodeling from sustained functional loss. The present evidence prioritizes that experiment; it does not provide a patient-risk predictor.

\hypertarget{conclusion}{%
\section{Conclusion}\label{conclusion}}

Public human hepatocyte data show an exploratory association between integrated stress responses and \emph{ABCB11} repression after adjustment for dose and specified liver-identity proxies. Comparator analyses extend the observed association to several other transporters, while sensitivity analyses characterize the influence of alternative adjustment models. External modules provide pathway-level consistency. These findings motivate multidonor protein/flux experiments and demonstrate why drug-safety evidence integration must retain source, exposure, readout and uncertainty.

\hypertarget{data-availability-and-ethics}{%
\section{Data availability and ethics}\label{data-availability-and-ethics}}

All analyses used publicly released transcriptomic, biochemical, structural, and regulatory data obtained from the repositories and supplementary materials cited in this article. No new participants were recruited, no interventions were performed, and no identifiable individual clinical records were processed. No new ethics approval or participant consent was sought for this secondary analysis.

\hypertarget{acknowledgments}{%
\section{Acknowledgments}\label{acknowledgments}}

We acknowledge the researchers and curators who made the DILImap and Open TG-GATEs datasets, TXG-MAPr resources, FDA DILIrank annotations, and experimental human BSEP structural coordinates publicly available. These contributions enabled the analyses and structural visualizations presented in this study.

\FloatBarrier
\bibliographystyle{IEEEtran}
\bibliography{references/references,references/ieee_control}

@IEEEtranBSTCTL{IEEEstyle:control,
  CTLuse_forced_etal = "yes",
  CTLmax_names_forced_etal = "6",
  CTLnames_show_etal = "1"
}

@article{aleo,
  author = {Aleo, Michael D. and Luo, Yi and Swiss, Rachel and Bonin, Paul D. and Potter, David M. and Will, Yvonne},
  title = {{Human drug-induced liver injury severity is highly associated with dual inhibition of liver mitochondrial function and bile salt export pump}},
  journal = {Hepatology},
  year = {2014},
  volume = {60},
  number = {3},
  pages = {1015--1022},
  doi = {10.1002/hep.27206}
}

@article{structure,
  author = {Liu, Hongtao and Irobalieva, Rossitza N. and Kowal, Julia and Ni, Dongchun and Nosol, Kamil and Bang-Sørensen, Rose and Lancien, Loïck and Stahlberg, Henning and Stieger, Bruno and Locher, Kaspar P.},
  title = {{Structural basis of bile salt extrusion and small-molecule inhibition in human BSEP}},
  journal = {Nature Communications},
  year = {2023},
  volume = {14},
  number = {1},
  pages = {7296},
  doi = {10.1038/s41467-023-43109-1}
}

@article{identity,
  author = {Dubois, Vanessa and Gheeraert, Céline and Vankrunkelsven, Wouter and Dubois-Chevalier, Julie and Dehondt, Hélène and Bobowski-Gerard, Marie and Vinod, Manjula and Zummo, Francesco Paolo and Güiza, Fabian and Ploton, Maheul and Dorchies, Emilie and Pineau, Laurent and Boulinguiez, Alexis and Vallez, Emmanuelle and Woitrain, Eloise and Baugé, Eric and Lalloyer, Fanny and Duhem, Christian and Rabhi, Nabil and van Kesteren, Ronald E and Chiang, Cheng-Ming and Lancel, Steve and Duez, Hélène and Annicotte, Jean-Sébastien and Paumelle, Réjane and Vanhorebeek, Ilse and Van den Berghe, Greet and Staels, Bart and Lefebvre, Philippe and Eeckhoute, Jérôme},
  title = {{Endoplasmic reticulum stress actively suppresses hepatic molecular identity in damaged liver}},
  journal = {Molecular Systems Biology},
  year = {2020},
  volume = {16},
  number = {5},
  pages = {e9156},
  doi = {10.15252/msb.20199156}
}

@article{isr,
  author = {Vlasveld, Matthijs and Callegaro, Giulia and Fisher, Ciarán and Eakins, Julie and Walker, Paul and Lok, Samantha and van Oost, Siddh and de Jong, Brechtje and Pellegrino-Coppola, Damiano and Burger, Gerhard and Wink, Steven and van de Water, Bob},
  title = {{The integrated stress response-related expression of CHOP due to mitochondrial toxicity is a warning sign for DILI liability}},
  journal = {Liver International},
  year = {2024},
  volume = {44},
  number = {3},
  pages = {760--775},
  doi = {10.1111/liv.15822}
}

@article{dilimap,
  author = {Bergen, Volker and Kodella, Konstantia and Srikrishnan, Sreenath and Barrandon, Ornella and Anderson, Sara and Rogers-Grazado, Max and Fowler, Casey and Beyene, Hirit and Robichaud, Nicole and Fulton, Timothy and Lapchyk, Nina and Cortes, Mauricio and Plugis, Nick and Goddeeris, Matthew and Zamanighomi, Mahdi},
  title = {{A large-scale human toxicogenomics resource for drug-induced liver injury prediction}},
  journal = {Nature Communications},
  year = {2025},
  volume = {16},
  number = {1},
  pages = {9860},
  doi = {10.1038/s41467-025-65690-3}
}

@article{txg,
  author = {Callegaro, Giulia and Kunnen, Steven J. and Trairatphisan, Panuwat and Grosdidier, Solène and Niemeijer, Marije and den Hollander, Wouter and Guney, Emre and Piñero Gonzalez, Janet and Furlong, Laura and Webster, Yue W. and Saez-Rodriguez, Julio and Sutherland, Jeffrey J. and Mollon, Jennifer and Stevens, James L. and van de Water, Bob},
  title = {{The human hepatocyte TXG-MAPr: gene co-expression network modules to support mechanism-based risk assessment}},
  journal = {Archives of Toxicology},
  year = {2021},
  volume = {95},
  number = {12},
  pages = {3745--3775},
  doi = {10.1007/s00204-021-03141-w}
}

@article{dilirank,
  author = {Olubamiwa, AyoOluwa O. and Qu, Yanyan and Connor, Skylar and Tong, Weida and Li, Dongying and Chen, Minjun},
  title = {{DILIrank 2.0: An updated and expanded database for drug-induced liver injury risk based on FDA labeling and a literature review}},
  journal = {Drug Discovery Today},
  year = {2025},
  volume = {30},
  number = {11},
  pages = {104485},
  doi = {10.1016/j.drudis.2025.104485}
}

@article{tggates,
  author = {Igarashi, Yoshinobu and Nakatsu, Noriyuki and Yamashita, Tomoya and Ono, Atsushi and Ohno, Yasuo and Urushidani, Tetsuro and Yamada, Hiroshi},
  title = {{Open TG-GATEs: a large-scale toxicogenomics database}},
  journal = {Nucleic Acids Research},
  year = {2014},
  volume = {43},
  number = {D1},
  pages = {D921--D927},
  doi = {10.1093/nar/gku955}
}

@article{deseq,
  author = {Love, Michael I and Huber, Wolfgang and Anders, Simon},
  title = {{Moderated estimation of fold change and dispersion for RNA-seq data with DESeq2}},
  journal = {Genome Biology},
  year = {2014},
  volume = {15},
  number = {12},
  pages = {550},
  doi = {10.1186/s13059-014-0550-8}
}

@article{seal,
  author = {Seal, Srijit and Williams, Dominic and Hosseini-Gerami, Layla and Mahale, Manas and Carpenter, Anne E. and Spjuth, Ola and Bender, Andreas},
  title = {{Improved Detection of Drug-Induced Liver Injury by Integrating Predicted In Vivo and In Vitro Data}},
  journal = {Chemical Research in Toxicology},
  year = {2024},
  volume = {37},
  number = {8},
  pages = {1290--1305},
  doi = {10.1021/acs.chemrestox.4c00015}
}

@article{bsep,
  author = {McLoughlin, Kevin S. and Jeong, Claire G. and Sweitzer, Thomas D. and Minnich, Amanda J. and Tse, Margaret J. and Bennion, Brian J. and Allen, Jonathan E. and Calad-Thomson, Stacie and Rush, Thomas S. and Brase, James M.},
  title = {{Machine Learning Models to Predict Inhibition of the Bile Salt Export Pump}},
  journal = {Journal of Chemical Information and Modeling},
  year = {2021},
  volume = {61},
  number = {2},
  pages = {587--602},
  doi = {10.1021/acs.jcim.0c00950}
}

@article{mito,
  author = {Hemmerich, Jennifer and Troger, Florentina and Füzi, Barbara and F.Ecker, Gerhard},
  title = {{Using Machine Learning Methods and Structural Alerts for Prediction of Mitochondrial Toxicity}},
  journal = {Molecular Informatics},
  year = {2020},
  volume = {39},
  number = {5},
  pages = {2000005},
  doi = {10.1002/minf.202000005}
}

@article{bsep2,
  author = {AbdulHameed, Mohamed Diwan M. and Liu, Ruifeng and Wallqvist, Anders},
  title = {{Using a Graph Convolutional Neural Network Model to Identify Bile Salt Export Pump Inhibitors}},
  journal = {ACS Omega},
  year = {2023},
  volume = {8},
  number = {24},
  pages = {21853--21861},
  doi = {10.1021/acsomega.3c01583}
}

@article{drugcentral,
  author = {Avram, Sorin and Bologa, Cristian G and Holmes, Jayme and Bocci, Giovanni and Wilson, Thomas B and Nguyen, Dac-Trung and Curpan, Ramona and Halip, Liliana and Bora, Alina and Yang, Jeremy J and Knockel, Jeffrey and Sirimulla, Suman and Ursu, Oleg and Oprea, Tudor I},
  title = {{DrugCentral 2021 supports drug discovery and repositioning}},
  journal = {Nucleic Acids Research},
  year = {2020},
  volume = {49},
  number = {D1},
  pages = {D1160--D1169},
  doi = {10.1093/nar/gkaa997}
}

@article{tolcapone,
  author = {Longo, DM and Yang, Y and Watkins, PB and Howell, BA and Siler, SQ},
  title = {{Elucidating Differences in the Hepatotoxic Potential of Tolcapone and Entacapone With DILIsym ® , a Mechanistic Model of Drug-Induced Liver Injury}},
  journal = {CPT: Pharmacometrics \& Systems Pharmacology},
  year = {2016},
  volume = {5},
  number = {1},
  pages = {31--39},
  doi = {10.1002/psp4.12053}
}

@article{hallmark,
  author = {Liberzon, Arthur and Birger, Chet and Thorvaldsdottir, Helga and Ghandi, Mahmoud and Mesirov, Jill P. and Tamayo, Pablo},
  title = {{The Molecular Signatures Database (MSigDB) hallmark gene set collection}},
  journal = {Cell Systems},
  year = {2015},
  volume = {1},
  number = {6},
  pages = {417--425},
  doi = {10.1016/j.cels.2015.12.004}
}
\end{document}


\maketitle
\hypertarget{supplementary-information}{%
\section*{Supplementary information}\label{supplementary-information}}

\hypertarget{s1-analysis-chronology-and-interpretation}{%
\subsection{S1 Analysis chronology and interpretation}\label{s1-analysis-chronology-and-interpretation}}

The initial assay specification preceded the four FDA-category associations. The transcriptomic extension was added after the sparse complete-case panel became apparent. Fixed RNA gene sets, within-compound estimation, exposure restrictions and influence checks were specified before fitting transcriptomic models. The broader published identity-panel projection and the external module analysis were subsequent exploratory additions. External modules were chosen from published gene membership before inspecting their response relationships. Source-unit restriction and descriptive identity/ubiquitous contrasts were added after the external and broad-panel results. Global-expression, alternative-stress, matched-random-set, comparator-outcome and LDH-proxy controls were later post hoc additions. All specifications are local, not registered preregistrations.

The two RNA transcript-endpoint p-values were Holm corrected as a stated pair. The original four assay tests form a separate, earlier Holm family. Follow-up models and gene-level comparisons are exploratory; their nominal p-values are not new confirmatory tests. Main inference emphasizes the compound-bootstrap intervals. Cluster-robust tests count absorbed compound intercepts; their uncertainty estimates may differ from percentile intervals in a restricted or heterogeneous panel.

\hypertarget{s2-complete-transcript-endpoint-estimates}{%
\subsection{S2 Complete transcript-endpoint estimates}\label{s2-complete-transcript-endpoint-estimates}}

Intervals use 2,000 compound-cluster resamples. Identity means the original four-gene score unless the broad score is specified. The exposure-restricted panel removes compounds with fewer than two remaining conditions; 87 source conditions met the exposure cutoff, but 80 entered within-compound inference. Similarly, 178 met the source MTD criterion, of which 177 entered repeated-dose inference. Cluster p-values are shown for completeness; the compound-bootstrap intervals remain the principal uncertainty summaries.

\begin{table}[htbp]\centering
\small\setlength{\tabcolsep}{4pt}\renewcommand{\arraystretch}{1.15}
\caption{Complete initial RNA transcript-endpoint estimates}
\begin{tabular}{@{}>{\raggedright\arraybackslash}p{1.55in}>{\raggedright\arraybackslash}p{0.68in}>{\raggedright\arraybackslash}p{0.84in}r >{\raggedright\arraybackslash}p{1.18in}r@{}}
\toprule
Model & Endpoint & Compounds / conditions & $\beta$ & 95\% interval & Cluster $p$ \\
\midrule
Unadjusted & ABCB11 & 50 / 197 & -0.673 & -0.829 to -0.524 & 9.8e-10 \\
Unadjusted & canalicular & 50 / 197 & -0.353 & -0.453 to -0.261 & 1.1e-07 \\
Dose & ABCB11 & 50 / 197 & -0.657 & -0.813 to -0.503 & 2.5e-09 \\
Dose & canalicular & 50 / 197 & -0.355 & -0.469 to -0.251 & 9.4e-07 \\
4-gene ID & ABCB11 & 50 / 197 & -0.302 & -0.477 to -0.103 & 0.0053 \\
4-gene ID & canalicular & 50 / 197 & -0.185 & -0.335 to -0.056 & 0.029 \\
Dose + 4-gene ID & ABCB11 & 50 / 197 & -0.321 & -0.494 to -0.127 & 0.0027 \\
Dose + 4-gene ID & canalicular & 50 / 197 & -0.200 & -0.349 to -0.074 & 0.019 \\
≤10× Cmax & ABCB11 & 27 / 80 & -0.726 & -0.943 to -0.509 & 1.1e-05 \\
≤10× Cmax & canalicular & 27 / 80 & -0.385 & -0.589 to -0.200 & 0.004 \\
≤10× + 4-gene ID & ABCB11 & 27 / 80 & -0.416 & -0.711 to 0.108 & 0.041 \\
≤10× + 4-gene ID & canalicular & 27 / 80 & -0.209 & -0.547 to 0.031 & 0.34 \\
Source MTD IC10 & ABCB11 & 45 / 177 & -0.676 & -0.843 to -0.514 & 8.5e-09 \\
Source MTD IC10 & canalicular & 45 / 177 & -0.351 & -0.463 to -0.253 & 6.8e-07 \\
\bottomrule\end{tabular}\end{table}

\hypertarget{s3-broad-identity-and-external-module-sensitivities}{%
\subsection{S3 Broad identity and external module sensitivities}\label{s3-broad-identity-and-external-module-sensitivities}}

The 436-gene LIVER-ID and 443-gene ubiquitous panels use exact case-normalized symbol projection of the published mouse panels, with complete source-response coverage and endpoint/ISR overlaps removed uniformly. Source identity-marker membership was independent of observed drug-response correlations. The external PHH:2 score represents a broad 743-gene module and does not constitute direct ABCB11 transcript replication.

\begin{table}[htbp]\centering
\small\setlength{\tabcolsep}{4pt}\renewcommand{\arraystretch}{1.15}
\caption{Broad identity sensitivity estimates}
\begin{tabular}{@{}>{\raggedright\arraybackslash}p{1.55in}>{\raggedright\arraybackslash}p{0.68in}>{\raggedright\arraybackslash}p{0.84in}r >{\raggedright\arraybackslash}p{1.18in}r@{}}
\toprule
Model & Endpoint & Compounds / conditions & $\beta$ & 95\% interval & Cluster $p$ \\
\midrule
436-gene ID & ABCB11 & 50 / 197 & -0.276 & -0.405 to -0.116 & 0.0013 \\
Dose + 436-gene ID & ABCB11 & 50 / 197 & -0.288 & -0.416 to -0.127 & 0.00086 \\
≤10× + 436-gene ID & ABCB11 & 27 / 80 & -0.372 & -0.556 to -0.011 & 0.00023 \\
436-gene ID & canalicular & 50 / 197 & -0.138 & -0.233 to -0.065 & 0.006 \\
Dose + 436-gene ID & canalicular & 50 / 197 & -0.153 & -0.252 to -0.077 & 0.0036 \\
≤10× + 436-gene ID & canalicular & 27 / 80 & -0.135 & -0.402 to -0.030 & 0.19 \\
\bottomrule\end{tabular}\end{table}

\begin{table}[htbp]\centering
\small\setlength{\tabcolsep}{4pt}\renewcommand{\arraystretch}{1.15}
\caption{External module estimates for 80 perturbagens}
\begin{tabular}{@{}>{\raggedright\arraybackslash}p{1.55in}>{\raggedright\arraybackslash}p{0.68in}>{\raggedright\arraybackslash}p{0.84in}r >{\raggedright\arraybackslash}p{1.18in}r@{}}
\toprule
Model & Endpoint & Compounds / conditions & $\beta$ & 95\% interval & Cluster $p$ \\
\midrule
Unadjusted & PHH:2 & 80 / 240 & -0.660 & -0.799 to -0.543 & 4.5e-13 \\
Dose & PHH:2 & 80 / 240 & -0.489 & -0.626 to -0.357 & 9.5e-08 \\
Identity & PHH:2 & 80 / 240 & -0.625 & -0.800 to -0.489 & 1.4e-09 \\
Dose + identity & PHH:2 & 80 / 240 & -0.476 & -0.621 to -0.351 & 6.4e-08 \\
\bottomrule\end{tabular}\end{table}

\begin{table}[htbp]\centering
\small\setlength{\tabcolsep}{4pt}\renewcommand{\arraystretch}{1.15}
\caption{External module estimates restricted to 69 micromolar-annotated perturbagens}
\begin{tabular}{@{}>{\raggedright\arraybackslash}p{1.55in}>{\raggedright\arraybackslash}p{0.68in}>{\raggedright\arraybackslash}p{0.84in}r >{\raggedright\arraybackslash}p{1.18in}r@{}}
\toprule
Model & Endpoint & Compounds / conditions & $\beta$ & 95\% interval & Cluster $p$ \\
\midrule
µM only, unadjusted & PHH:2 & 69 / 207 & -0.670 & -0.814 to -0.544 & 1.5e-11 \\
µM only + dose & PHH:2 & 69 / 207 & -0.506 & -0.660 to -0.355 & 6.4e-07 \\
µM only + identity & PHH:2 & 69 / 207 & -0.658 & -0.841 to -0.508 & 3.4e-09 \\
µM only + dose + identity & PHH:2 & 69 / 207 & -0.507 & -0.654 to -0.380 & 5.6e-08 \\
\bottomrule\end{tabular}\end{table}

\hypertarget{s4-initial-biochemical-associations-and-source-ascertainment}{%
\subsection{S4 Initial biochemical associations and source ascertainment}\label{s4-initial-biochemical-associations-and-source-ascertainment}}

The following four initial comparisons use exact conditional odds ratios and confidence limits, with Holm adjustment across the four tests. The complete model panel had 40 most/no parents, below the prespecified prediction gate of 50. No joint classifier performance is claimed. The assay-source thresholds, identity harmonization and missingness handling are described in the main manuscript.

\begin{table}[htbp]\centering
\small\setlength{\tabcolsep}{4pt}\renewcommand{\arraystretch}{1.15}
\caption{Initial biochemical associations with FDA concern categories}
\begin{tabular}{@{}>{\raggedright\arraybackslash}p{0.42in}>{\raggedright\arraybackslash}p{0.87in}r >{\raggedright\arraybackslash}p{0.66in}>{\raggedright\arraybackslash}p{0.73in}r >{\raggedright\arraybackslash}p{1.12in}r@{}}
\toprule
Assay & Contrast & $N$ & Most active & Control active & OR & Exact 95\% interval & Holm $p$ \\
\midrule
BSEP & most vs no & 73 & 15 / 35 & 13 / 38 & 1.435 & 0.504 to 4.140 & 1.000 \\
BSEP & most vs less & 108 & 15 / 35 & 25 / 73 & 1.435 & 0.577 to 3.546 & 1.000 \\
Mito & most vs no & 121 & 9 / 59 & 3 / 62 & 3.505 & 0.817 to 21.212 & 0.283 \\
Mito & most vs less & 171 & 9 / 59 & 13 / 112 & 1.368 & 0.481 to 3.735 & 1.000 \\
\bottomrule\end{tabular}\end{table}

Source-sensitivity estimates are exploratory. The two BSEP sources agree on 161 of 170 overlapping FDA parents, yet category associations change with compound coverage. Odds ratios across different panels are not directly comparable causal effects, and their visual difference is not a formal test of source-effect heterogeneity.

\subsection{S5 Additional specificity controls}

The fixed ISR score was compared with 5,000 expression-matched and 5,000 expression- and response-variance-matched six-gene sets after within-compound centering and dose/436-gene identity adjustment. Absolute partial correlation was the comparison statistic. The ISR correlation ($r=-0.422$) was equaled or exceeded by 1 and 16 sets, respectively. With the stated $(1+k)/5001$ calculation, the empirical fractions were approximately 0.0004 and 0.0034. These are random-set benchmarks, not compound-bootstrap significance levels or proof of pathway exclusivity.

Global-expression controls were computed from 16,133 eligible genes, with named ISR, endpoint, comparator-outcome and broad-identity genes excluded. The ROS, apoptosis and ISR-disjoint unfolded-protein-response scores contained 48, 145 and 105 genes, respectively. The dose/identity model jointly including median and absolute expression change plus ROS and apoptosis scores used 197 conditions from 50 compounds and yielded $\beta=-0.198$ (95\% compound-bootstrap interval, $-0.342$ to $-0.039$).

Only 38 retained RNA conditions had exact LDH-screen dose matches, and just one compound had repeated exact matches. The sensitivity analysis therefore used 38 exact matches and 114 within-range interpolations, yielding 152 conditions from 39 compounds; no dose extrapolation was used. Adding this proxy to all global-expression and alternative-stress controls yielded $\beta=-0.190$ ($-0.335$ to $0.009$). The interval includes zero. LDH screening was performed in parallel source experiments and does not provide viability measurements in the RNA-seq wells. All additional controls were post hoc, and their intervals are pointwise.

\FloatBarrier\clearpage
\hypertarget{s5-complete-panel-and-source-sensitivity-figures}{%
\subsection{S6 Complete-panel and source-sensitivity figures}\label{s5-complete-panel-and-source-sensitivity-figures}}

\begin{figure}[H]
\centering
\includegraphics[width=\linewidth]{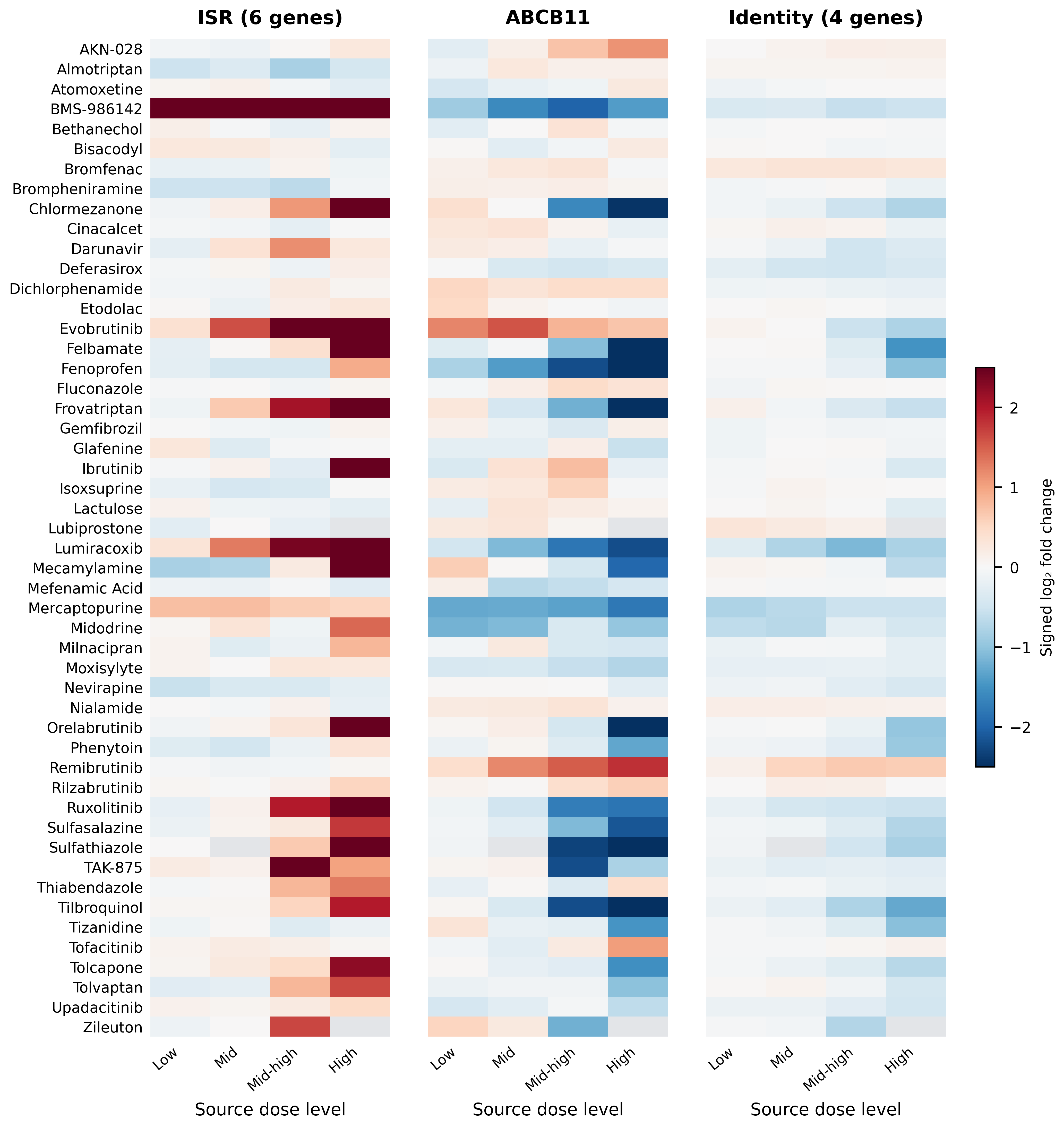}
\caption{All 50 retained RNA compound dose series. Missing or excluded source conditions are gray. Colors display signed log2 fold changes relative to source controls and saturate at ±2.5; saturation of the color scale does not imply truncation of the source measurements. Dose level is source-relative and is not a common absolute concentration across compounds.}
\end{figure}

\begin{figure}[H]
\centering
\includegraphics[width=\linewidth]{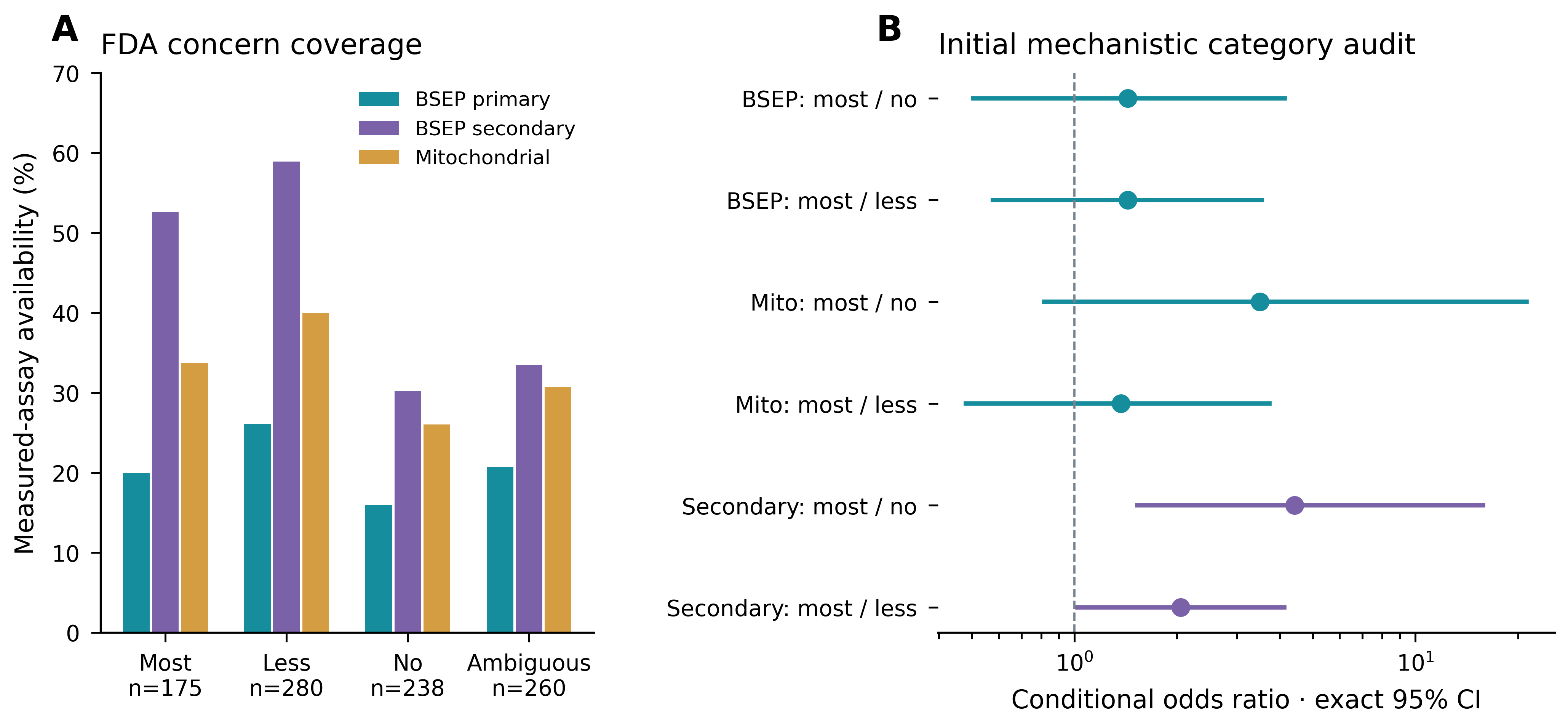}
\caption{Assay availability and initial FDA concern associations. The primary comparisons and overlapping secondary-source sensitivities are shown with exact conditional intervals. Missing assays remain missing. FDA concern labels describe regulatory/literature categories, not clinical incidence.}
\end{figure}

\begin{figure}[H]
\centering
\includegraphics[width=\linewidth]{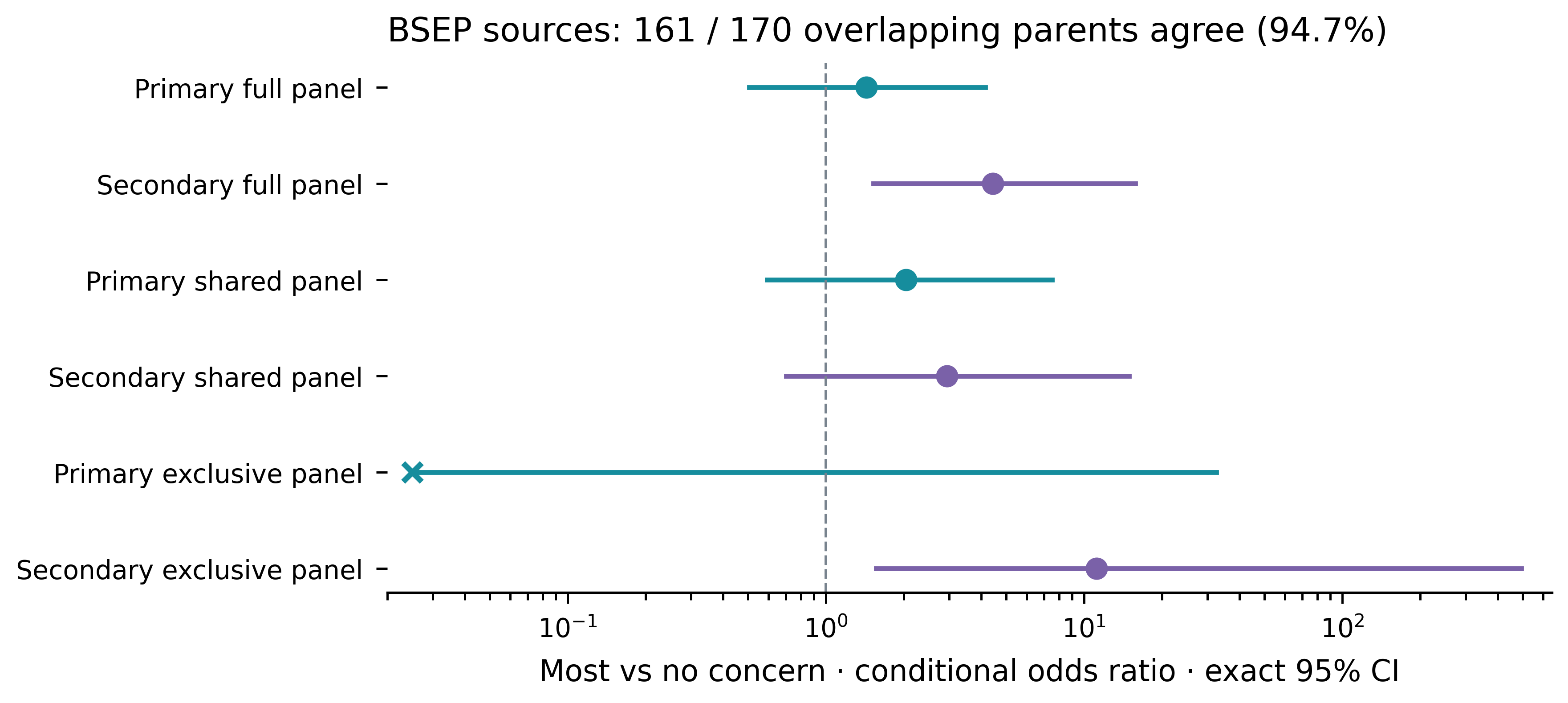}
\caption{BSEP source sensitivity for most versus no concern. The primary and secondary full panels, the exact shared panel, and source-exclusive panels differ in coverage. The x mark indicates the primary-exclusive zero odds-ratio estimate; its lower interval extends to zero and is truncated at the plotted log-axis boundary. The secondary-exclusive interval is extremely wide.}
\end{figure}

\FloatBarrier\clearpage
\hypertarget{s6-complementary-three-dimensional-protein-context}{%
\subsection{S7 Complementary three-dimensional protein context}\label{s6-complementary-three-dimensional-protein-context}}

Supplementary Figure S4 presents complementary ribbon and molecular-surface views of the experimental human BSEP--glibenclamide structure used in Figure 1 of the main manuscript \href{https://doi.org/10.1038/s41467-023-43109-1}{[2]}. Both views use rigid rotations and translations of deposited coordinates, with no conformational interpolation, missing-domain reconstruction or docking. NBD1 remains unresolved.

\begin{figure}[H]
\centering
\includegraphics[width=\linewidth]{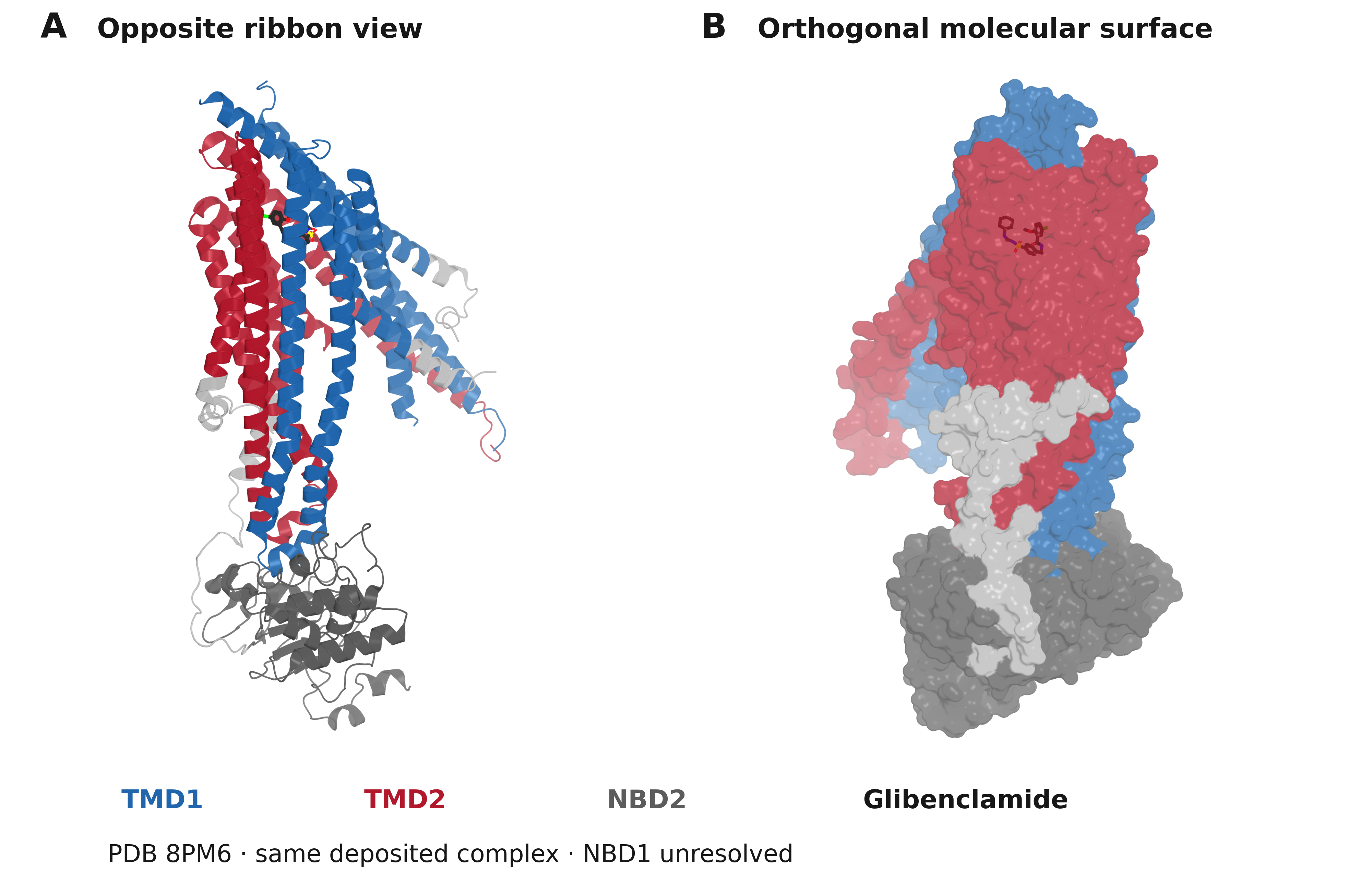}
\caption{Complementary views of experimental human BSEP, PDB 8PM6. A: opposite ribbon view. B: orthogonal semitransparent molecular surface with deposited glibenclamide. TMD1 is blue, TMD2 red and resolved NBD2 gray; ligand carbons are charcoal and heteroatoms colored by element. Both views use the same atomic coordinates; NBD1 is unresolved.}
\end{figure}

\hypertarget{s7-focused-prospective-validation}{%
\subsection{S8 Focused prospective validation}\label{s7-focused-prospective-validation}}

A follow-up should use multiple independent primary human hepatocyte donors with matched vehicle controls and repeated experimental exposures. Pair RNA measurement with BSEP protein abundance and canalicular localization, bile-salt transport flux, and viability. Measure intracellular and unbound exposure where feasible. Include direct BSEP inhibitor reference compounds, low-response comparators, and the three descriptive examples. Sample before and after 24 hours so transcriptional change, protein loss and functional effects can be temporally ordered. A stress-pathway perturbation or rescue condition can test dependence on the ISR; a simple expression correlation cannot establish that dependence. Prespecify the primary functional endpoint, drug/donor clustering, exposure range and multiplicity approach before new experiments.

This is a proposed validation strategy. None of these prospective experiments, transport measurements or pathway-rescue results has been performed in the present study.

\renewcommand\refname{References}
  
\FloatBarrier